\documentclass[12pt]{emulateapj}
\usepackage{natbib}
\usepackage{rotating}
\usepackage{mathptmx}

\newcommand{\kms}{$\mbox{km~s}^{-1}$}

\newcommand{\lta}{\raisebox{-0.6ex}{$\,\stackrel
{\raisebox{-.2ex}{$\textstyle <$}}{\sim}\,$}}
   
\newcommand{\msol}{\mbox{M$_{\odot}$}}

\newcommand{\twospec}[2] {
 \begin{center}
    \begin{minipage}[t]{0.45\textwidth}
      \includegraphics[angle=270,scale=0.4]{#1}
    \end{minipage}
    \hfill
    \begin{minipage}[t]{0.45\textwidth}
      \includegraphics[angle=270,scale=0.4]{#2}
    \end{minipage}
  \end{center}
}

\shorttitle{36 GHz methanol emission in NGC\,253}
\shortauthors{}

\begin{document}

\title{Detection of 36~GHz class I methanol maser emission towards NGC\,253}

\author{Simon P. Ellingsen\altaffilmark{1}, Xi~Chen\altaffilmark{2,3}, Hai-Hua Qiao\altaffilmark{2,4}, 
Willem Baan\altaffilmark{2}, Tao An\altaffilmark{2,3}, Juan Li\altaffilmark{2,3}, Shari L. Breen\altaffilmark{5}}

\altaffiltext{1} {School of Physical Sciences, University of
Tasmania, Hobart, Tasmania, Australia; Simon.Ellingsen@utas.edu.au}

\altaffiltext{2} {Shanghai Astronomical Observatory, Chinese Academy
of Sciences, Shanghai 200030, China}

\altaffiltext{3} {Key Laboratory of Radio Astronomy, Chinese Academy
of Sciences, Nanjing, JiangSu 210008, China}

\altaffiltext{4} {University of Chinese Academy of Sciences, 19A
Yuquanlu, Beijing 100049, China}

\altaffiltext{5} {CSIRO Astronomy and Space Science, Australia
Telescope National Facility, PO Box 76, Epping, NSW 1710, Australia}

\begin{abstract}

We have used the Australia Telescope Compact Array (ATCA) to search for emission from the $4_{-1} \rightarrow 3_{0}E$ transition of methanol (36.2~GHz) towards the center of the nearby starburst galaxy NGC\,253.  Two regions of emission were detected, offset from the nucleus along the same position angle as the inner spiral arms.  The emission is largely unresolved on a scale of 5\arcsec, has a full-width half maximum (FWHM) line width of $<$ 30~\kms, and an isotropic luminosity orders of magnitude larger than that observed in any Galactic star formation regions.  These characteristics suggest that the 36.2~GHz methanol emission is most likely a maser, although observations with higher angular and spectral resolution are required to confirm this.  If it is a maser this represents the first detection of a class~I methanol maser outside the Milky Way.  The 36.2~GHz methanol emission in NGC\,253 has more than an order of magnitude higher isotropic luminosity than the widespread emission recently detected towards the center of the Milky Way.  If emission from this transition scales with nuclear star formation rate then it may be detectable in the central regions of many starburst galaxies.  Detection of methanol emission in ultra-luminous infra-red galaxies (ULIRGs) would open up a new tool for testing for variations in fundamental constants (in particular the proton-to-electron mass ratio) on cosmological scales.

\end{abstract}


\section{Introduction}

Methanol is a commonly observed species in interstellar gas, and its rich microwave and millimetre spectrum has made it a powerful tool for studying high-mass star formation through its numerous masing transitions \citep[e.g.][]{Ellingsen+12}, while thermal methanol emission is often observed towards hot molecular cores \citep[e.g.][]{vanderTak+00a}.  

It has recently been discovered that observations of rotational transitions of methanol provide particularly sensitive tests for variations in the proton-to-electron mass ratio ($\mu$) \citep{Jansen+11,Levshakov+11}.  Although readily observable in the local universe, the only detections of methanol emission beyond the Milky Way are a small number of masers from the 6.7~GHz transition towards the Large Magellanic Cloud \citep[LMC;][]{Green+08} and M31 \citep{Sjouwerman+10}, one 12.2~GHz maser in the LMC \citep{Ellingsen+10}, absorption in the 6.7~GHz transition towards the center of NGC\,3079 \citep{Impellizzeri+08} and thermal emission towards the nearby galaxies NGC\,253, IC342, Maffei 2, NGC\,6946, NGC\,4945 and M82 \citep{Henkel+87,Henkel+90,Huttemeister+97,Martin+06}.  Methanol absorption has been detected in a number of transitions towards the lensing galaxy ($z=$ 0.89) in the PKS B\,1830-211 gravitational lens system \citep{Ellingsen+12b,Bagdonaite+13,Muller+14} and these observations have been used to place tight constraints on variations in $\mu$ in this system \citep{Bagdonaite+13}.  However, strong lensing systems such as PKS\,B1830-211 are rare, so an alternative approach for detecting methanol at high-redshifts is desirable.

Some active galaxies show strong maser emission from either the 1667 MHz OH or the 22~GHz water transitions.  These masers are in some cases more than a million times more luminous than typical galactic masers observed in star formation regions or late-type stars and for this reason are referred to as megamasers.  OH megamasers are observed towards the nuclear regions of ULIRGS, systems undergoing merger activity \citep[e.g.][]{Baan+82}, while water masers are observed in accretion disks and jets towards low-luminosity AGN, typically Seyfert 2 galaxies \citep[e.g.][]{Miyoshi+95}.  A number of unsuccessful searches for methanol megamasers have been undertaken \citep{Ellingsen+94,Phillips+98a,Darling+03}, these have all focused on the 6.7~GHz ($5_{1} \rightarrow 6_{0}A^{+}$) transition of methanol, the strongest and most common transition in galactic star formation regions.  The targets for these searches were primarily known OH and water megamaser systems.

Methanol maser transitions are empirically divided into two classes based on their environment and pumping mechanism.  Class I masers are collisionally pumped where outflows or other low-energy shocks interact with dense molecular gas.  In general, multiple discrete maser sites are observed towards a single star formation region, distributed on scales of around 1~pc \citep{Kurtz+04,Voronkov+14}.  Class II masers are radiatively pumped and are closely associated with infrared sources, OH and water masers.  Generally only one or two sites are observed in a given star formation region and the masers are distributed on scales an order of magnitude smaller than the class I masers.  The class~II methanol masers are exclusively associated with high-mass star formation regions \citep{Breen+13b}, whereas class~I masers have also been observed associated with lower-mass young stars \citep{Kalenskii+10} and recently have also been detected associated with supernova remnants \citep{Pihlstrom+14}.  To date, there have been no reported searches for extragalactic class~I masers.

Recently, \citet{Yusef-Zadeh+13} detected widespread emission from the $4_{-1} \rightarrow 3_{0}E$  class~I maser transition of methanol (which we hereafter call the 36.2~GHz methanol transition), in the inner region of the Milky Way.  Observations with the Very Large Array detected more than 350 separated sites in a 160 $\times$ 43 pc region, with an integrated luminosity over the whole region in excess of 5600 Jy \kms.  The 36.2~GHz class~I methanol maser transition was first detected towards galactic star formation regions \citep{Haschick+89b} and is one of the most common and strongest class~I methanol masers \citep{Voronkov+14}.  \citet{Yusef-Zadeh+13} suggest that the large number of 36.2~GHz methanol masers in the central molecular zone (CMZ) of the Milky Way is due to enhanced methanol abundance in this region produced by photodesorption of methanol from cold dust by cosmic rays.  It is likely that similar mechanisms operate in other galaxies and furthermore, galaxies with enhanced star formation in their central regions may exhibit 36.2 GHz emission over larger volumes which may be readily detectable.  Interestingly, more than 20 years ago \citet{Sobolev93} suggested that the 36.2~GHz class I methanol transition provided the best prospect for producing methanol megamasers.  

In this paper we report a search for 36.2~GHz methanol emission from the central region of  NGC\,253, a nearby spiral galaxy which has significant starburst activity towards its center \citep[see][and references therein]{Sakamoto+11}.  NGC\,253 has been studied in detail at a wide range of wavelengths, including radio \citep[e.g.][]{Ulvestad+97}, millimetre \citep[e.g.][]{Bolatto+13}, infrared \citep[e.g.][]{Dale+09}, optical \citep[e.g.][]{Dalcanton+09} and x-ray \citep[e.g.][]{Strickland+00}.  Large amounts of molecular gas have been detected towards the central region of the galaxy \citep{Mauersberger+96}, with more than 25 different molecular species having been observed \citep[e.g.][]{Martin+06}, including the first detection of methanol emission beyond the Milky Way \citep{Henkel+87}.  We have adopted a distance estimate of 3.4~Mpc for NGC\,253 in this work \citep{Dalcanton+09}.

\section{Observations}

The observations were made using the ATCA on 2014 March 29 (project code C2879).  The array was in the H168 configuration (baseline lengths between 61 and 192 m) and the synthesised beam width for the observations at 36~GHz was approximately 8.0\arcsec $\times$ 4.2\arcsec\/ The Compact Array Broadband Backend \citep[CABB;][]{Wilson+11} was configured with 2 x 2.048 GHz bands centered on frequencies of 35.3 and 37.3 GHz respectively.  Each of the bands was divided into 2048 spectral channels each of 1 MHz bandwidth, corresponding to a spectral resolution of 9.9~\kms\/ at 36.2 GHz for uniform weighting of the correlation function (a channel width of 8.2~\kms).  The 2.048 GHz bands cover the rest frequencies of the $4_{-1} \rightarrow 3_{0}E$ and $7_{-2} \rightarrow 8_{-1}E$ transitions of methanol, for which we adopted rest frequencies of 36.169265 and 37.703700 GHz respectively \citep{Muller+04}.

The data were reduced with {\sc miriad} using the standard techniques for ATCA observations.  Amplitude calibration was with respect to Uranus and PKS\,B1921-293 was observed as the bandpass calibrator.  The data were corrected for atmospheric opacity and the absolute flux density calibration is estimated to be accurate to 30\%.  The observing strategy interleaved 10 minutes on NGC\,253 (pointing center $\alpha = 00^{\mbox{h}}47^{\mbox{m}}33.10^{\mbox{s}}$ ; $\delta = -25\arcdeg17\arcmin18.0\arcsec$ (J2000) ) with 2 minute observations of a nearby phase calibrator (0116-219) before and after the target source.  The total duration of the on-source time for NGC\,253 was 30 minutes.  The data from the two bands were combined during imaging to yield a total bandwidth of $\sim$3.9 GHz.  Continuum emission from center of NGC\,253 was then used to produce a model for self-calibration.  After self-calibration continuum subtraction was undertaken prior to imaging the 36.2 and 37.7~GHz methanol maser transitions at full spectral resolution.  The resulting RMS noise in a single 1~MHz (8.2~\kms) spectral channel for the 36.2~GHz methanol transition was \lta 0.8 mJy beam$^{-1}$.

Additional observations were undertaken in a director's time allocation on 2014 May 18.  The array was in the 1.5D configuration (baseline lengths between 107 and 1469 m).  We obtained a further 30 minutes on-source on NGC\,253, but with very limited hour-angle coverage (approximately 45 minutes).  The amplitude and bandpass calibration were performed in the same manner as for the March observations.

\section{Results} \label{sec:results}

\begin{figure*}
 \begin{center}
    \begin{minipage}[t]{1.0\textwidth}
      \includegraphics[scale=0.45]{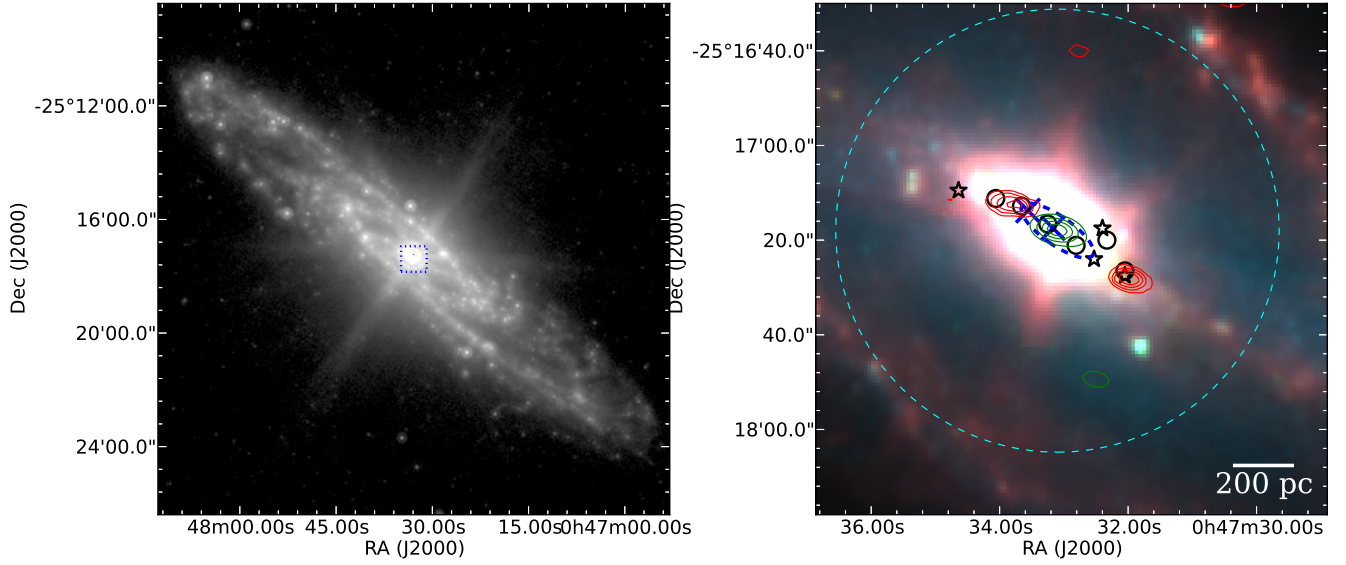}
    \end{minipage}
  \end{center}  
\caption{{\em Left:} {\em Spitzer} 24~$\mu$m image of NGC\,253 \citep{Dale+09}.  The blue box marks the region shown in the right-hand panel.  {\em Right:} 36~GHz continuum emission (green contours at 20, 40, 60 and 80\% of 95 mJy beam$^{-1}$) and 36.2~GHz methanol emission (red contours 20, 40, 60 and 80\% of 310 mJy \kms\/ beam$^{-1}$).  The dashed blue ellipse marks the half-maximum intensity of the central molecular zone of NGC\,253 \citep{Sakamoto+11}.  The background image is from {\em Spitzer} IRAC observations with blue, green and red from the 3.6, 4.5 and 8.0~$\mu$m bands respectively \citep{Dale+09}.  The blue crosses mark the location of water masers detected by \citet{Henkel+04}, the black stars mark of supersshells identified by \citet{Sakamoto+06} and \citet{Bolatto+13} and the black circles NH$_3$ cores \citep{Lebron+11}.  The cyan dashed line shows the size of the primary beam of the ATCA observations} \label{fig:ngc253}
\end{figure*}

The continuum emission from the central region of NGC\,253 was detected as a point source in our observations with an integrated intensity of 0.11 Jy.  Molecular emission within the central molecular zone (CMZ) of NGC\,253 spans a velocity range from $\sim 100$--400 \kms\/ \citep{Sakamoto+11} and we imaged a velocity range covering barycentric velocities in the range -150 -- 670~\kms\/ for the 36.2~GHz methanol transition  at 8.2~\kms\/ velocity resolution.  Figure~\ref{fig:ngc253} (right) shows the both the 36~GHz continuum emission (green contours) and the integrated 36.2~GHz methanol emission (red contours) with a {\em Spitzer} IRAC 3-color image as the background \citep{Dale+09}.  The 36.2~GHz methanol emission was detected in both observing epochs, however for the May 18 observations, the purely east-west baseline configuration and limited hour angle range mean that it cannot be used for imaging and the data presented here is only from the March 29 observations.  For the May 18 observations the 36.2~GHz emission is observed on all baselines shorter than 25~k$\lambda$, but not on the longer baselines.  As the line emission is detected with relatively low signal to noise in a 30 minute observation it is not possible to determine from our data if the non-detection on longer baselines is because the emission is resolved, or if it is due to atmospheric decorrelation.

The continuum emission is clearly associated with the nucleus of NGC\,253, while the methanol emission is offset.  At the resolution of our observations we detect two regions of 36.2~GHz methanol emission (Table~\ref{tab:results}), significantly offset from the nucleus, but lying along the same position angle as the inner spiral arms traced by the mid-infrared emission.  

The observations of the 37.7~GHz methanol transition found no emission with a 5$\sigma$ limit of approximately 3 mJy beam$^{-1}$ in a single 1~MHz spectral channel.  The 37.7~GHz transition of methanol is observed as a relatively rare and weak class~II maser in Galactic star formation regions \citep[e.g.][]{Ellingsen+11a}, so its non-detection here is expected.  This transition was only included because it could be observed simultaneously with the 36.2~GHz transition and we do not discuss it further here.

The methanol emission offset from the nucleus towards the north-east (hereafter Meth NE) is associated with two ammonia cores \citep{Lebron+11} and is close to one of two water masers detected towards NGC\,253 by \citet{Henkel+04} the locations of which are marked in Fig.~\ref{fig:ngc253} with blue crosses.  Meth NE is offset from the center of NGC\,253 by 11.1\arcsec, which for an assume distance of 3.4~Mpc corresponds to 180~pc.  The other methanol emission towards the south-west (hereafter Meth SW) is offset from the center by 18.4\arcsec\/ (300~pc) and is associated with an ammonia core and is close to the cite of a supershell identified by \citet{Bolatto+13}.

\begin{table*}
  \begin{center} 
  \caption{36.2~GHz emission towards NGC\,253} \label{tab:results}
  \begin{tabular}{lllrrrr}
  {\bf Name} & {\bf Right Ascension} & {\bf Declination} & {\bf Peak Flux} & {\bf Integrated Flux} & {\bf Peak Velocity} & {\bf Velocity range} \\ 
                      & {\bf ($^h$ $^m$ $^s$)} & {\bf ($^{\circ}$ \arcmin\  \arcsec)} & {\bf (mJy)} & {\bf (mJy \kms)} & {\bf (\kms)} & {\bf (\kms)} \\ \hline
    Meth. NE & 00:47:33.8 & -25:17:12 & 10.5 & $310\pm50$ &  194 & 88 -- 211 \\ 
    Meth. SW & 00:47:32.0 & -25:17:28 & 8.7   & $310\pm50$ &  350 & 285 -- 350 \\  \hline
    Continuum & 00:47:33.1 & -25:17:18 & 93    & 110 & \\ \hline
  \end{tabular}
  \end{center} 
\end{table*}

\section{Discussion}

\subsection{The nature of the 36 GHz methanol emission}

\begin{figure*}
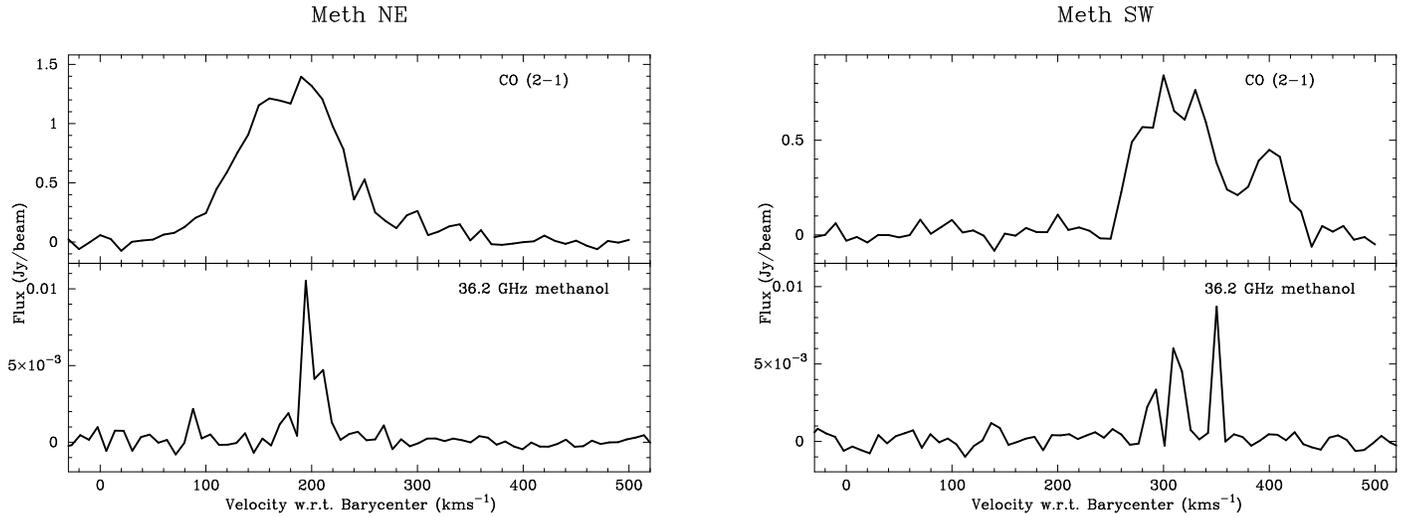

 \twospec{meth_ne}{meth_sw}
 \caption{Comparison of the 36.2~GHz methanol emission from the currently observations with the CO(2--1) emission from the same region \citet{Sakamoto+11}} \label{fig:co_meth}
\end{figure*}

These observations represent the first detection of methanol emission from the $4_{-1} \rightarrow 3_{0}E$ from a source beyond the Milky Way.   A key question to address is whether the 36.2~GHz methanol emission sites in NGC\,253 are produced by maser or thermal processes.  Figure~\ref{fig:co_meth} compares the spectra of the 36.2~GHz methanol emission to that from CO(2--1) towards the same location \citep{Sakamoto+11}.  It is clear from Fig.~\ref{fig:co_meth} that the velocity of the methanol emission is close to peak of the CO(2-1) from the same direction, but covers a significantly smaller velocity range.  Previous observations with similar angular resolution have detected methanol emission towards the central regions of NGC\,253 with full width to half maximum (FWHM) line widths of around 100~\kms \citep{Henkel+87,Martin+06}.  In contrast, for the 36.2~GHz methanol emission in Fig.~\ref{fig:co_meth} the lines are narrow, with widths of the order of the spectral resolution of our observations ($\sim$10~\kms).

The isotropic luminosity of Milky Way class~I methanol masers associated with star formation regions is typically less than 500 Jy \kms\/ kpc$^2$, while the integrated 36.2~GHz methanol emission from the Milky Way CMZ is three orders of magnitude greater.  We measure an integrated intensity of 0.62 Jy \kms\/ for the 36.2 GHz methanol emission towards the central region of NGC\,253 (Table~\ref{tab:results}), which for a distance of 3.4~Mpc corresponds to an isotropic luminosity of $7.2 \times 10^{6}$ Jy~\kms\/ kpc$^2$.  At a distance of 8.5~kpc this would correspond to an integrated intensity of approximately 10$^5$~Jy~\kms\/.  Hence the 36.2~GHz methanol emission towards the center of NGC\,253 is a factor of more than 17 greater than the emission in the Milky Way CMZ, much, perhaps most of which is known to be due to maser sources.  

Figure~\ref{fig:ngc253} shows the two methanol emission regions are compact and essentially unresolved at the resolution of these observations.  Using the Rayleigh-Jeans approximation, the factor to convert an unresolved source of this angular size from a flux density to the temperature scale is 40 K Jy$^{-1}$.  \citet{Martin+06} detected multiple thermal transitions of methanol in the 2~mm band toward the center of NGC 253. Assuming optically thin emission for the observed methanol transitions \citeauthor{Martin+06} used rotation diagram analysis to derive a column density and rotation temperature of $8.3 \times 10^{14}$ cm$^{-2}$ and 11.6 K, respectively. Using these values we can estimate the integrated intensity to expect in the 36.2~GHz transition if the emission is thermal.  We calculate an integrated intensity of 0.8 K \kms\/ for thermal emission from the 36.2~GHz transition, however, we observe an integrated intensity of 25 K \kms\/ (corresponding to 0.62 Jy \kms\/ see Table~\ref{tab:results}). So the integrated emission we observe from the 36.2~GHz methanol transition is about 30 times greater than predicted if it were due to thermal emission similar to that previously observed from methanol in NGC\,253.

In combination, the high isotropic luminosity, well in-excess of the expectations for thermal emission and the narrow line width of the 36.2~GHz methanol emission in NGC\,253 is highly suggestive that it is maser emission.  However, additional observations with higher spectral and angular resolution are required to verify that the 36.2~GHz methanol emission in NGC\,253 is a maser.

\subsection{Comparison with the Milky Way CMZ methanol emission}

The CMZ in NGC\,253 is observed to have an extent (to half-maximum intensity) of about 300x100 pc \citep[indicated by a blue dashed ellipse in Figure~\ref{fig:ngc253};][]{Sakamoto+11} and the two regions of 36.2~GHz methanol emission identified here lie beyond that.  This is within the region where H$_2$CO absorption is observed \citep{Baan+97} and close the the outer extent of emission from dense gas tracers such as HCN, HCO$^+$ and NH$_3$ \citep{Knudsen+07,Lebron+11}.  The mechanism proposed by \citet{Yusef-Zadeh+13} to explain Milky Way CMZ 36.2~GHz methanol masers requires the presence of cold dust (to form or host methanol-rich ices) and an enhanced cosmic ray flux (compared to the galactic disc). \citet{Yusef-Zadeh+13} found that high cosmic ray intensities desorb methanol from cold dust grains, but if the intensity is too high they are rapidly destroyed once they are released into the gas phase.  They explain the low density of 36.2 GHz methanol emission within the circumnuclear ring of the Milky Way as perhaps being due to this.  The absence of any strong 36.2~GHz methanol emission in the inner 100~pc in NGC\,253 may be because within that region gas-phase methanol is rapidly destroyed.  This would imply that the flux of cosmic rays and/or energetic photons are at extreme levels over a much larger volume than in the Milky Way, which is plausible, given the high supernova rate estimated for this region \citep{Lenc+06}.  Alternatively, there is evidence that the edges of the inner molecular disk in NGC\,253 are affected by the nuclear outflow and the 36.2~GHz methanol emission may be being produced by that interaction.  Methanol is released from dust grains in slow shocks, and this has previously been suggested as the mechanism responsible for the presence of easily dissociated molecules in the central region of NGC\,253 \citep{Lebron+11}, and the close association of both sites of 36.2~GHz methanol emission with NH$_3$ cores supports this hypothesis (Fig.~\ref{fig:ngc253}).  A third possibility is that the 36.2~GHz methanol emission sites correspond to regions where there is a large number of very young high-mass star formation regions within a small volume and the Meth SW emission is close to a supershell \citep{Bolatto+13}.  However, in terms of the intensity of the dust continuum and mid-infrared emission the two methanol emission sites are much weaker than regions closer to the NGC\,253 nucleus, which is inconsistent with them being locations of unusual star formation activity within the central regions of NGC\,253.  So the observed methanol emission sites likely represent regions where there is an appropriate balance between the cold dust where methanol is produced through grain surface reactions and the presence of cosmic rays or slow shocks to release it into the gas phase.    

\subsection{Prospects for 36.2~GHz methanol megamasers}

The star formation rate in NGC\,253 has been estimated from the infrared luminosity to be 3.6 \msol yr$^{-1}$ \citep{Strickland+04}, significantly higher than recent estimates for the Milky Way of 0.68 -- 1.45 \msol yr$^{-1}$ determined from {\em Spitzer} observations \citep{Robitaille+10}.   However, the starburst activity in NGC\,253 is relatively modest and more distant, interacting galaxies exhibit star formation rates up to two or three orders of magnitude greater.  The detected 36.2~GHz methanol emission towards the center of NGC\,253 has an isotropic luminosity four orders of magnitude stronger than that observed from typical class~I masers in galactic star formation regions, so it does not classify as a methanol megamaser.  However, it is more than an order of magnitude stronger than the integrated 36.2~GHz methanol emission from the Milky Way CMZ, which suggests that emission from this transition may scale with star formation rate.  Additional observations of other nearby starburst systems are clearly required to determine if this is the case and to assess the specific factors which govern the luminosity of the 36.2~GHz methanol emission. The detection of 36.2~GHz methanol masers in the Milky Way CMZ and NGC\,253 suggests that galaxies with large amounts of cold dust and high star formation rates within their central regions (e.g. ULIRGs) may host large volumes conducive to enhanced 36.2~GHz methanol emission.

As outlined in the introduction, one motivation for searching for methanol emission at cosmological distances is that it is unusually sensitive to variations in the proton-to-electron mass ratio $\mu$.  The sensitivity coefficient $K_{\mu}$ for the 36.2~GHz $4_{-1} \rightarrow 3_{0}E$ transition is approximately 10 \citep{Jansen+11,Levshakov+11}, an order of magnitude larger than a pure rotational transition. Different methanol transitions have different sensitivities to variations in $\mu$ and as discussed by \citet{Ellingsen+11b}, comparison between methanol maser transitions is able to provide constraints with far fewer sources of systematic uncertainty than comparisons between different molecular species.  The 84.5~GHz $5_{-1} \rightarrow 4_{0}E$ transition \citep[$K_{\mu} \sim  3.5$;][]{Levshakov+11} and the 229~GHz $8_{-1} \rightarrow 7_{0}E$ transition are in the same family as the 36.2~GHz transition and are observed as class~I masers in galactic star formation regions.  If the 36.2~GHz $4_{-1} \rightarrow 3_{0}E$ transition is detected as a megamaser at cosmological distances, then sensitive searches for the 84.5 and 229~GHz transitions should be undertaken to determine whether they can also be observed in these systems.

\section{Conclusions}

We have made the first search for the 36.2~GHz $4_{-1} \rightarrow 3_{0}E$ methanol transition towards the nearby starburst Galaxy NGC\,253.  We detect two regions of emission, close to, but offset from the nuclear region.  The line width and luminosity compared to previous observations of thermal methanol emission in NGC\,253 suggests that the 36.2~GHz emission regions are likely masers, although further data are required to confirm this.  If the 36.2~GHz emission is from masers then this represents the first detection of class~I methanol masers outside the Milky Way and the first detection of any methanol maser at distances greater than 1~Mpc.  If emission from this transition scales with star formation rate then it may be possible to detect it at cosmological distances and use it to test for variations in the fundamental constants.

\section*{Acknowledgements}

Xi Chen thanks to the supports from the National Natural Science Foundation of China (11133008 and 11273043), the Strategic Priority Research Program of the Chinese Academy of Sciences (CAS;  Grant No. XDA04060701), Key Laboratory for Radio Astronomy, CAS. Shari Breen is the recipient of an Australian Research Council DECRA Fellowship (project number DE130101270).   The Australia Telescope Compact Array is part of the Australia Telescope National Facility which is funded by the Commonwealth of Australia for operation as a National Facility managed by CSIRO. 

\end{document}